\documentclass[prl, twocolumn,amsmath,amssymb,showpacs,superscriptaddress]{revtex4-1}

\usepackage{graphicx} 

\begin{document}

\title{Strange nonchaotic stars}

\author{John F. Lindner}
\affiliation{Department of Physics and Astronomy, University of Hawai`i at M\=anoa Honolulu, Hawai`i 96822, USA}
\affiliation{Physics Department, The College of Wooster, Wooster, Ohio 44691, USA}

\author{Vivek Kohar}
\affiliation{Department of Physics and Astronomy, University of Hawai`i at M\=anoa Honolulu, Hawai`i 96822, USA}

\author{Behnam Kia}
\affiliation{Department of Physics and Astronomy, University of Hawai`i at M\=anoa Honolulu, Hawai`i 96822, USA}

\author{Michael Hippke}
\affiliation{Institute for Data Analysis, Luiter Stra{\ss}e 21b, 47506 Neukirchen-Vluyn, Germany}

\author{John G. Learned}
\affiliation{Department of Physics and Astronomy, University of Hawai`i at M\=anoa Honolulu, Hawai`i 96822, USA}

\author{William L. Ditto}
\affiliation{Department of Physics and Astronomy, University of Hawai`i at M\=anoa Honolulu, Hawai`i 96822, USA}

\date{\today}

\begin{abstract}
The unprecedented light curves of the \textit{Kepler} space telescope document how the brightness of some stars pulsates at primary and secondary frequencies whose ratios are near the golden mean, the most irrational number. A nonlinear dynamical system driven by an irrational ratio of frequencies generically exhibits a strange but nonchaotic attractor. For \textit{Kepler}'s ``golden" stars, we present evidence of the first observation of strange nonchaotic dynamics in nature outside the laboratory. This discovery could aid the classification and detailed modeling of variable stars.
\end{abstract}

\pacs{05.45.Tp, 05.45.Df, 97.10.Sj, 95.75.Wx}

\maketitle 


As a byproduct of its vastly successful search for exoplanets~\cite{Borucki,Chaplin,Finkbeiner,Koch,Jenkins}, the \textit{Kepler} spacecraft has revolutionized stellar photometry by its high precision tracking of the brightness of 150~000 stars nearly continuously for four years. Both the quantity and quality of the \textit{Kepler} data enable new stellar discoveries. The \textit{Kepler} stars include cosmic distance markers (or standard candles) such as a Cepheid and 41 RR Lyrae variable stars. While some of these stars pulsate with a single frequency, \textit{Kepler} observations confirm that others pulsate with multiple frequencies. Several of these stars, including the RRc Lyrae star KIC 5520878, pulsate with two principal frequencies, which are nearly in the golden ratio~\cite{Livio}. As the most irrational number, the golden ratio can have significant dynamical consequences. For example, by the KAM theorem, dynamics with two frequencies in the golden ratio maximally resist perturbations~\cite{Hilborn}. Furthermore, nonlinear systems driven by two incommensurate frequencies, forming an irrational ratio~\cite{Cubero}, and especially the golden ratio, exhibit distinctive dynamics that exist between order and chaos. Strange (because fractal) \textit{nonchaotic} attractors characterize this distinctive dynamics~\cite{Grebogi, Feudel}. Here we report evidence for strange nonchaotic behavior in the pulsations of stars like KIC 5520878, making this golden star the prototype of a new class. Strange nonchaotic attractors have only been seen in laboratory experiments but never before in nature. The observation of stellar strange nonchaotic dynamics provides a new window into variable stars that can improve their classification and refine the physical modeling of their interiors~\cite{Regev}.

\begin{figure}[hb!] 
	\includegraphics[width=0.76\linewidth]{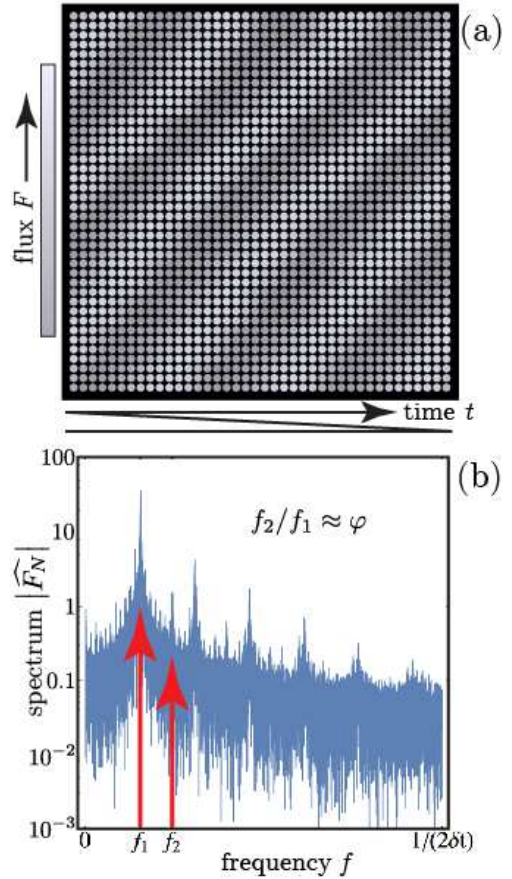} 
	\caption{Spectral content. (a) Raster pattern of disks with brightness proportional to the KIC 5520878 stellar flux at intervals of $\delta t \approx 1766~\text{s} \approx 0.5~\text{h}$. (b) Fourier transform magnitude of stellar flux sampled at intervals $\delta t$ has primary and secondary frequencies (red arrows) at $f_1 \approx 1/(0.266~\text{d})$ and $f_2 \approx 1/(0.169~\text{d})$, where $f_2 / f_1 \approx 1.58 \approx \varphi$ is the golden ratio.}
	\label{SpectralContent}
\end{figure}

\begin{figure*}[hbt]
	\includegraphics[width=0.755\linewidth]{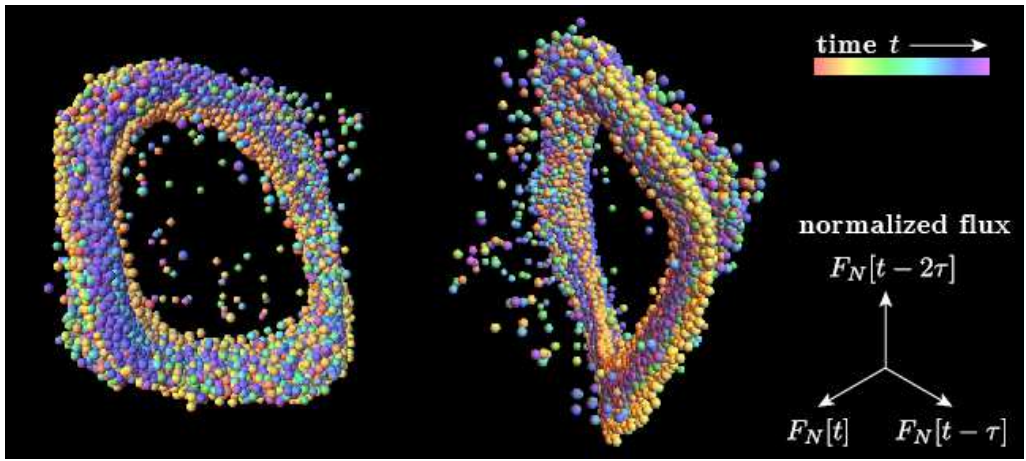} 
	\caption{Attractor reconstruction. Two views of a three-dimensional plot of normalized flux $F_N$ at $t = n \delta t$ successively delayed by $\tau = 2 \delta t$. This delay coordinate embedding ``unfolds" the time series into a warped torus, suggesting two-frequency nonlinear dynamics. Equal-sized spheres locate data. Rainbow colors code time, from red to violet. Flux triplets straddling data gaps appear far from torus. }
	\label{AttractorReconstruction}
\end{figure*}

Periodically driven damped nonlinear systems can exhibit complex dynamics characterized by strange chaotic attractors, where strange refers to the fractal geometry \textit{of} the attractor and chaotic refers to the exponential sensitivity of orbits \textit{on} the attractor. Quasiperiodically driven systems forced by incommensurate frequencies are natural extensions of periodically driven ones and are phenomenologically richer. In addition to periodic or quasiperiodic motion, they can exhibit chaotic or nonchaotic motion on strange attractors. Although quasiperiodic forcing is not necessary for strange nonchaotic dynamics, the first experiment to demonstrate a strange nonchaotic attractor~\cite{Ditto} involved the buckling of a magnetoelastic ribbon driven quasiperiodically by two incommensurate frequencies in the golden ratio $f_1/f_2  = \varphi =(1+ \sqrt{5})/2 \approx 1.62$, which is the irrational number least well approximated by rational numbers, as its continued fraction expansion $\varphi = 1+1/ \varphi=1+1/(1+1/(1+ \cdots))$ is all 1s. Here we apply the time series analysis techniques of the ribbon experiment to analyze the dynamics of several \textit{Kepler} multi-frequency variable stars, including the RRc Lyrae star KIC 5520878, a blue-white star $16~000$ light years from Earth in the constellation Lyra whose brightness varies as in Fig.~\ref{SpectralContent}(a).

\begin{figure*}
	\includegraphics[width=0.73\linewidth]{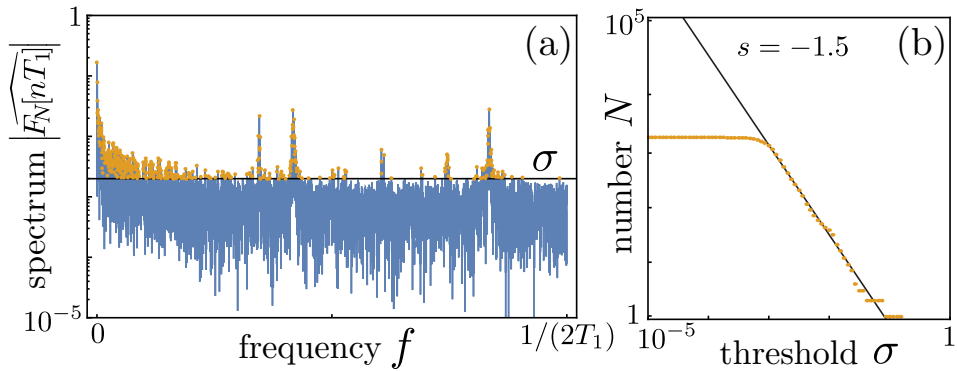} 
	\caption{Spectral scaling. (a) Fourier transform magnitude of stellar flux strobed at the primary period $T_1$ has a wide distribution of peaks. Peaks above threshold $\sigma$ (horizontal line) are highlighted (gold). (b) Log-log plot of number of super-threshold peaks $N$ versus the threshold $\sigma$ has a linear regime (in leg below knee of curve) that indicates power law scaling and fractal structure. Spectral exponent slope $-2<s<-1$ indicates strange nonchaotic dynamics. Deviation for low thresholds arises from light curve's finite temporal resolution. }
	\label{SpectralDistribution}
\end{figure*}

\begin{figure}[ht] 
	\includegraphics[width=0.73\linewidth]{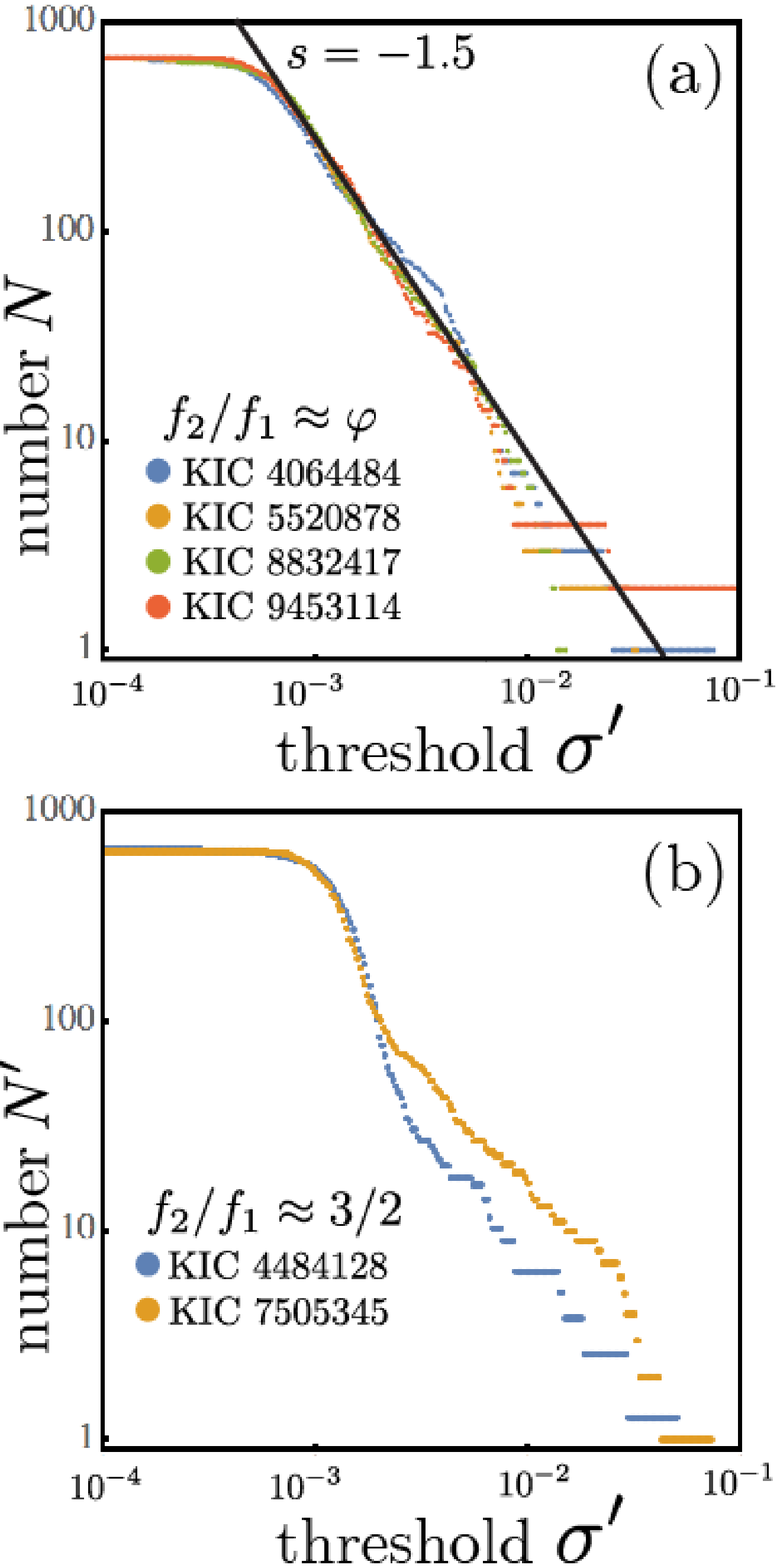} 
	\caption{Other stars. (a) Spectral scalings of the four \textit{Kepler} golden stars all exhibit power law scaling, over 2 decades of thresholds, and nearly collapse onto each other when the axes are proportionally scaled (as denoted by primes). (b) By contrast, spectral scalings of two \textit{Kepler} non-golden stars with primary and secondary frequency ratios near $3/2$ do not exhibit power law scaling but are in fact concave or multi-scaled.}
	\label{OtherStars}
\end{figure}

Most classical Cepheids and RR Lyrae are single-mode periodic and radial pulsators. For many years, the only known exception was long-term Blazhko light curve modulation found among some RR Lyrae~\cite{Blazhko}. Recently, better quality photometry has uncovered various multi-mode pulsations within these variable stars, possibly from non-radial modes. A few dozen such stars have been documented~\cite{Moskalik} for mysterious behavior involving frequency ratios near 0.62, apparently without connection to the golden ratio $\varphi \approx 1/0.62$ or the possibility of strange nonchaotic behavior. We refer to the former stars as \textit{golden stars} and to the latter stars as strange \textit{nonchaotic stars}, and we recognize that these classes may overlap.

We downloaded the \textit{Kepler} Input Catalogue 5520878 data from the Mikulski Archive for Space Telescopes. The light curve, or flux $F$ (in detector electrons per second) versus time $t$ (in days), was sampled every $\delta t = 1766$ seconds or about once each half hour, for most of four years, including over 5000 primary periods. To keep its solar cells in sunlight, the \textit{Kepler} spacecraft rotated $90^\circ$ every quarter solar orbit causing the star to illuminate different pixels of its charge-coupled device. Different pixels have different responses and appear to have contributed to shifts and skews in the data segments. In addition, events such as cosmic ray strikes, monthly science data downlinks, and safe modes caused both small and large gaps in the data. The best detrending practice for \textit{planet} searches~\cite{Kipping} cannot be employed for variable stars, such as KIC 5520878, because each pulsation is intrinsically different. Instead, we manually processed the raw time series $F$ by detrending and rescaling the data segments to zero mean and unit variance to obtain a refined or normalized time series $F_N$.

A natural strategy to test the flux time series for strange nonchaotic behavior is to estimate both the dimension of the underlying attractor and the largest Lyapunov exponent of the orbits. A fractional dimension would imply fractal geometric scaling and strangeness, while a positive Lyapunov exponent would imply the divergence of nearby orbits and chaos~\cite{Hilborn}. Although these metrics are difficult to estimate~\cite{Plachy2014}, we investigated both.

For the fractal geometry~\cite{SM}, we calculated the \textit{information} dimension, which typically delivers the most accurate dimension estimates for short noisy time series~\cite{Farmer} by quantifying how the information needed to specify a set of points scales with the area that contains them. To estimate it, we partitioned a section of the attractor into many tiny boxes, computed the fraction of points in each box, and calculated the corresponding information or entropy. The logarithmic scaling of information with slope $d \sim 1.8$ in the range $1<d<2$ is consistent with some fractal structure. (In contrast, for pure quasiperiodic motion, the section is a sinusoidal curve of dimension $d=1$, and for completely noisy data, the section is a uniform area of dimension $d=2$.) 

For the largest Lyapunov exponent~\cite{SM}, we implemented both the Wolf \cite{Wolf} and Rosenstein \cite{Rosenstein} algorithms and found the latter more accurate and stable when tested on known time series. Beginning with the Fig.~\ref{AttractorReconstruction} attractor reconstruction, we found the nearest neighbors of all points not too close together. We then computed the average separation of all neighboring points as a function of iteration. A linear increase in the logarithm of the average separation versus iteration number would have indicated a positive largest Lyapunov exponent. However, we did not observe such a linearity, which is consistent with a zero largest Lyapunov exponent \cite{Rosenstein}. 

Since these two obvious metrics are difficult to estimate, we focused instead on the spectral scaling of the dynamics, which is a more reliable measure that has traditionally been used to identify strange nonchaotic motion~\cite{Ditto, Ding1989, Romeiras}. We analyzed the spectral scaling in two independent data pipelines with comparable results. One pipeline used \textit{Mathematica} 10 software while the other used MATLAB, Period04, and custom C++ software.

From a Fourier transform $\hat F_N [f]$ of the normalized time series $F_N [t]$, we identified the two most significant frequencies in the light curve, as shown in Fig.~\ref{SpectralContent}(b), where red arrows indicate primary and secondary frequencies at $f_1 \approx 1/(0.266~\text{d})$ and $f_2 \approx 1/(0.169~\text{d})$, corresponding to primary and secondary periods of $T_1 \approx 6.41~\text{h}$ and $T_2 \approx 4.05~\text{h}$, where $f_1 / f_2   =T_2/T_1 \approx 1.58 \approx \varphi$ is nearly the incommensurate golden ratio, in agreement with previous work~\cite{Moskalik, Hippke}. (Other spectral peaks correspond to harmonics of $f_1$ and linear combinations of $f_1$ and $f_2$.) We next plotted three-dimensional delay coordinates $\{F_N [t],F_N [t-\tau],F_N [t-2 \tau] \}$ at times $t = n \delta t$ and delay $\tau=2\delta t$ to reconstruct the time series' underlying attractor~\cite{Takens}. The resulting Fig.~\ref{AttractorReconstruction} torus is consistent with quasiperiodic forcing of the dynamics. Points far from the torus straddle gaps in the data, and the warp of the torus reflects the nonlinearity of the light curve \cite{SM}. 

One way to take a Poincar\'{e} section of a state space flow is to intercept it with a plane, but a second way is to sample or strobe it at a fixed frequency. As in the original strange nonchaotic experiment \cite{Ditto}, we strobed the data at the primary period $T_1$ and wrapped it by the secondary period $T_2$ to produce the section. To qualitatively check the section dynamics \cite{SM}, we tracked the normalized flux differences $\Delta F_N=F_N [(n_1+n) T_1 ]-F_N [(n_2+n)T_1 ]$ of two points $n_1$ and $n_2$ as a function of iteration $n$. For many steps, this difference is small, briefly increases to large, and rapidly returns to small, a characteristic nonchaotic behavior. (In contrast, for strange \textit{chaotic} attractors, the difference would start small, increase to large, and thereafter remain uncorrelated, without the repeated rapid close returns.)

To quantify the section geometry, we examined the magnitude of the discrete Fourier transform of the section (or strobed) data, as in Fig.~\ref{SpectralDistribution}(a), and recorded the number of peaks $N$ above a threshold $\sigma$. The power law scaling of this spectral distribution $N=N_0 \sigma^s$, over 2 decades of thresholds and 3 decades of number, with an exponent $s \approx -1.5$ in the range $-2<s<-1$, as in the Fig.~\ref{SpectralDistribution}(b) log-log plot, is a classic signature of strange nonchaotic dynamics~\cite{Grebogi, Feudel}. The power law means that the distribution of peaks is scale free, with a range of large and small peaks, presumably reflecting the fractal nature of the underlying strange attractor~\cite{Takens, Hunt}. The deviation at low thresholds results from the light curve's finite frequency resolution.

The cumulative case for strange nonchaotic dynamics includes the strobed or section spectrum, the spectral exponent, the distinctive orbit separation behavior, and to a lesser extent the section dimension and Lyapunov exponent. The identification is challenging because of inevitable noise, limited and imperfect (despite unprecedented) data, and the smallness of the secondary frequency amplitude compared to the primary frequency amplitude (as suggested by the Fig.~\ref{SpectralContent}(b) spectral peak heights). Quasi-periodic dynamics naturally flows on a two-dimensional torus, with the two frequencies driving the toroidal and poloidal motions; quasi-periodic forcing of a nonlinear system naturally inhabits a wrinkled torus with fractal cross section, but in this case the wrinkles may be small (or the strangeness higher dimensional), and standard analysis might miss them.

We tested our time series analysis by analyzing a variety of null hypotheses that generated surrogate data sets of artificial light curves of flux versus time, including both nonparametric and parametric models. Nonparametric models included a classic phase randomization of the original time series, which was the inverse Fourier transform of the phase randomized Fourier transform of the light curve~\cite{Theiler}. The strobed section and spectral exponent easily discriminated between the surrogate and original data. 

Parametric models included an ideal quasiperiodic curve having either the KIC 5520878 near-golden frequency ratio or the ideal golden frequency ratio, with or without noise added to the fluxes or the times. The spectral distributions of these artificial data sets were not strange nonchaotic, demonstrating that the KIC 5520878 pulsations are not simply noisy quasiperiodic. To check the effects of the many small and the few large data gaps, we linearly or sinusoidally interpolated to close them. This produced some artifacts in the section plot, but still produced a spectral exponent in the strange nonchaotic regime. Finally, we introduced the exact experimental gaps into the ideal quasiperiodic curve. For limited ranges of additive noise and thresholds, this did produce spectral power law scaling, but the strobed spectra remained qualitatively distinct, and strange nonchaotic dynamics remained the best explanation.

We expanded this analysis to five other multi-frequency variable stars in \textit{Kepler}'s field of view. The three additional golden RRc Lyrae stars KIC 4064484, KIC 8832417 and KIC 9453114, whose frequency ratios well approximate the golden mean, all exhibit signatures of strange nonchaotic dynamics, and their spectral scalings nearly collapse onto each other when the axes are proportionally scaled, as in Fig.~\ref{OtherStars}(a). In contrast, two non-golden RRab Lyrae stars KIC 4484128 and KIC 7505345~\cite{Benko}, whose frequency ratios well approximate the simple fraction $3/2$, exhibit qualitatively different dynamics, as in Fig.~\ref{OtherStars}(b). The possibility of other golden stars with strange nonchaotic behavior, discovered by ongoing large sky surveys like OGLE~\cite{Soszynski} or ASAS~\cite{Pojmanski}, promises to further classify the RR Lyrae and Cepheid variable stars.

Our model-independent nonlinear analysis of the light curves of such variable stars is complementary to detailed nonlinear hydrodynamic models of the stars themselves~\cite{Plachy2013}. The strange nonchaotic signatures of variable stars like KIC 5520878 may elucidate phenomena like the Eddington valve mechanism~\cite{Eddington} thought to underlie their pulsations: variations in the opacity of the star might quasiperiodically modulate the normal hydrostatic balance between pressure outward and gravity inward, thereby generating light curves somewhere between order and chaos that the best models will need to reproduce. 

Strange nonchaotic attractors have been observed in laboratory experiments involving magnetoelastic ribbons~\cite{Ditto}, electrochemical cells~\cite{Ruiz}, electronic circuits~\cite{Zhou}, and a neon glow discharge~\cite{Ding1997}, but never before in non-experiments in nature. The pulsating star KIC 5520878 may be the first strange nonchaotic dynamical system observed in the wild.

\pagebreak

We gratefully acknowledge the entire Kepler team, whose outstanding work has made possible our results, as well as support from the Office of Naval Research under Grant No. N00014-12-1-0026 and STTR grant No. N00014-14-C-0033. J.F.L. thanks The College of Wooster for making possible his sabbatical at the University of Hawai'i at M\=anoa.


\end{document}